# Zero External Magnetic Field Quantum Standard of Resistance at the $10^{-9}$ Level

D. K. Patel,[1,*] K. M. Fijalkowski,[2,3] M. Kruskopf,[1] N. Liu,[2,3] M. Götz,[1] E. Pesel,[1] M. Jaime,[1] M. Klement,[2,3] S. Schreyeck,[2,3] K. Brunner,[2,3] C. Gould,[2,3] L. W. Molenkamp,[2,3] and H. Scherer[1,†]

[1] *Physikalisch-Technische Bundesanstalt (PTB), Bundesallee 100, 38116 Braunschweig, Germany*
[2] *Faculty for Physics and Astronomy (EP3), Universität Würzburg, Am Hubland, D-97074, Würzburg, Germany*
[3] *Institute for Topological Insulators (ITI), Am Hubland, D-97074, Würzburg, Germany*

∗ email: dinesh.patel@ptb.de
† email: hansjoerg.scherer@ptb.de

## ABSTRACT

The quantum anomalous Hall effect holds promise as a disruptive innovation in condensed matter physics and metrology, as it gives access to Hall resistance quantization in terms of the von-Klitzing constant $R_K = h/e^2$ at zero external magnetic field. In this work, we study the accuracy of Hall resistance quantization in a device based on the magnetic topological insulator material $(V,Bi,Sb)_2Te_3$. We show that the relative deviation of the Hall resistance from $R_K$ at zero external magnetic field is $(4.4 \pm 8.7)$ nΩ/Ω when extrapolated to zero measurement current, and $(8.6 \pm 6.7)$ nΩ/Ω when extrapolated to zero longitudinal resistivity (each with combined standard uncertainty, $k = 1$), which sets a new benchmark for the quantization accuracy in topological matter. This precision and accuracy at the nΩ/Ω level (or $10^{-9}$ of relative uncertainty) achieve the thresholds for relevant metrological applications and establish a zero external magnetic field quantum standard of resistance — an important step towards the integration of quantum-based voltage and resistance standards into a single universal quantum electrical reference.

## INTRODUCTION

Since the revision of the International System of Units (SI) in 2019 [1], quantum Hall resistance standards (QHRS) and Josephson voltage standards (JVS) provide primary realizations of the units of electrical resistance (ohm) and voltage (volt), respectively. Common material systems conventionally used in resistance metrology for implementations of QHRS devices are GaAs/AlGaAs heterostructures [2] and graphene [3–5], which require a superconducting solenoid to generate a strong magnetic field for the realization of the quantum Hall effect (QHE) [6]. This reliance on high magnetic fields limits the practicality of such QHRS and makes it challenging to realize them together with JVS (which becomes inoperable in even small magnetic fields) in a single cryogenic apparatus [7, 8].

The recent breakthrough discovery of the quantum anomalous Hall effect (QAHE) [9, 10] in magnetically doped topological insulators (TIs) such as V- or Cr-doped $(Bi,Sb)_2Te_3$ allows for the realization of Hall resistance quantization in terms of the von-Klitzing constant $R_K$ without a permanent external magnetic field [8–14]. This enables a combined realization of the resistance and voltage standards. However, accessing the QAHE still remains demanding because of two requirements on the measurement conditions: very low temperatures of typically below 50 mK [10, 11] and relatively low bias currents of typically well below 1 μA [12, 13, 15, 16]. The primary factor constraining the performance appears to stem from limited insulating properties of the bulk/surface states, which − when activated − electrically short the edge channels that carry the dissipationless currents [17–19], and thus impair the quantization.

Metrological measurements in V- and Cr-doped $(Bi,Sb)_2Te_3$ already quite early enabled the realization of the QAHE and the demonstration of quantization at the $10^{-7}$ to $10^{-6}$ level of relative accuracy and precision [12–14, 20]. These results, while an important step in improving the QAHE, are still orders of magnitude away from even early stages of integer quantum Hall devices based on epitaxial graphene, for which accuracies at the parts-per-billion ($10^{-9}$) level were reported [3]. Recently, a level of $10^{-8}$ has been achieved in a TI device applying an external magnetic field of some 200 mT [21]. In terms of measurement uncertainty, the nΩ/Ω (or $10^{-9}$) level represents the threshold for metrological state-of-the art applications and services at the performance level typically supplied by national metrology institutes [22]. The realization of a true zero external magnetic field quantum standard of resistance operating at the accuracy level of order of $10^{-9}$ is therefore not only an important step forward in establishing new types of QHRS, but also in developing an integrated universal and primary electrical reference (QHRS together with JVS) which necessitates operation in the absence of any external magnetic field.

In this work, we demonstrate that the relative deviation of the Hall resistance from $R_K$ at zero external magnetic field in our device is $(4.4 \pm 8.7)$ n$\Omega/\Omega$ when extrapolated to zero current, and $(8.6 \pm 6.7)$ n$\Omega/\Omega$ when extrapolated to zero longitudinal resistivity (each with combined standard uncertainty, $k = 1$). This realization of $R_K$ at the parts-per-billion level of measurement precision and accuracy marks the realization of an external magnetic field-free quantum standard of electrical resistance.

## DEVICE PREPARATION AND BASIC CHARACTERIZATION

A 9 nm thick layer of V-doped topological insulator $V_{0.1}(Bi_{0.2}Sb_{0.8})_{1.9}Te_3$ is grown using molecular beam epitaxy (MBE) on a hydrogen passivated Si(111) substrate, and covered by an in-situ grown 10 nm thick insulating Te capping layer [23]. Standard photolithography techniques are used to fabricate the Hall bar device (200 $\mu$m wide, 730 $\mu$m long). An optical microscope image of the device is shown in Fig. 1a. The device features nine electrodes, including an electrostatic gate, source/drain, and three pairs of Hall voltage contacts. The aspect ratio of the distance between two neighboring voltage contacts to the width of the Hall bar is 1:1. To pattern the ohmic contacts, the Te cap is locally removed using Ar milling, followed by electron beam evaporation of a 100 nm thick layer stack of AuGe/Ti/Au without breaking the high vacuum conditions. The device is fitted with a top gate consisting of 20 nm AlOx/HfOx and 100 nm Ti/Au, deposited using atomic layer deposition (ALD) and electron beam evaporation, respectively. The gate stack is given in Fig. 1b. In the final fabrication step, the sample is glued onto a chip carrier and electrically connected with wedge-bonded Au wires.

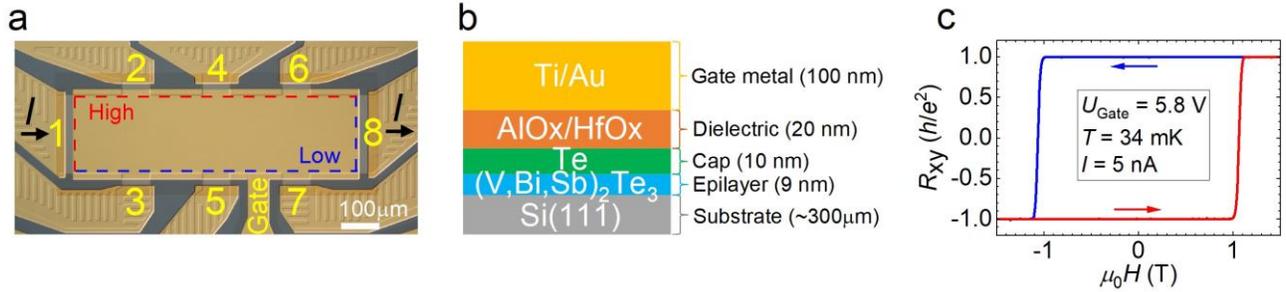

FIG. 1. **Device layout and basic characterization.** a) Optical Nomarski microscope image of the device, with overlaid contact labels (1 to 8), from which source and drain contacts are labeled as 1 and 8, voltage terminals as 2-7, and a top gate contact as "Gate". The red and blue dashed lines indicate the high and low potential edges of the device for magnetization pointing normal-to-plane. b) Schematic depiction of the layer structure of the device. c) Hall resistance $R_{xy}$ versus normal-to-plane magnetic field $H$. The Hall voltage is measured at contact pair 4-5 (during up and down sweep of the magnetic field) using standard low-frequency lock-in characterization techniques, at temperature $T = 34$ mK, current $I = 5$ nA, and applied top gate voltage $U_{Gate} = 5.8$ V.

All data in this paper are collected with current applied through terminals 1 and 8 and by measuring the voltages between various pairs of terminals from 2 to 7, at zero external magnetic field (with exception of the magnetic field sweep characterization measurement in Fig. 1c) after having applied the normal-to-plane magnetic field of $\mu_0 H = 1.5$ T to establish the magnetization direction. Measurements are done at base temperature of the dilution refrigerator (about 34 mK). The insulation resistance of the top gate is determined to be $R_{iso} = 6.9$ T$\Omega$ (see Fig. S1 in the Supplementary Information). The optimal gate voltage for the QAHE measurements that maximizes the insulating properties of the bulk, indicated by the minimum value of longitudinal resistance $R_{xx}$, is $U_{Gate} = 5.8$ V (see Fig. S2 in the Supplementary Information). The contact resistance is found to be of the order of 10 $\Omega$ for all current and voltage terminals on our device, fulfilling the requirements defined by the guidelines for reliable direct current (DC) measurements of the quantized Hall resistance in metrology [22].

Fig. 1c depicts a typical characterization measurement: the Hall resistance $R_{xy}$ measured during sweeps of the external normal-to-plane magnetic field, with an alternating current (AC) excitation of 5 nA applied at a frequency of 13.7 Hz and a gate voltage $U_{Gate} = 5.8$ V. The data is collected using standard lock-in techniques. The curve shows the typical ferromagnetic hysteresis loop behavior expected from the QAHE, with the Hall resistance switching between $+h/e^2$ and $-h/e^2$ depending on the magnetization direction in the film. Noteworthy, in contrast to conventional QHRS devices used in resistance metrology based on GaAs/AlGaAs heterostructures or graphene, requiring permanent external magnetic fields of significant strength corresponding to some 10 T of magnetic induction, our TI device only needs a relatively weak magnetic field of order of $\mu_0 H = 1$ T temporarily to be applied for establishing the magnetization. For the following operation as QHRS, the external field is then switched off, which enables the operation of the device integrated with superconducting circuits like JVS. This operation mode is also different from the approach in Ref. [21], which was based on the permanent application of a small magnetic field of some 200 mT, eliminating a need for a superconducting solenoid magnet.

PRECISION MEASUREMENTS

The term "precision measurements" refers to the transport measurements performed with a resistance bridge based on a cryogenic current comparator (CCC), in contrast to the basic characterization measurements performed with the lock-in setup described in the previous section. In this work, we use a state-of-the-art 14-bit cryogenic current comparator (CCC), designed and built in-house at PTB [24–26], which has about four times higher number of turns in the coils compared to the 12-bit CCC system routinely used for early QAHE measurements [12–14, 20, 21]. Fig. 2a shows a simplified schematic of the CCC resistance bridge.

A stable and calibrated 100 Ω resistor is used as reference. Details on its temporal stability and calibration are given in the Supplementary Information. A DC superconducting quantum interference device (SQUID) coupled to the CCC coils detects the net flux generated by the counter-propagating currents $I_1$ and $I_2$ in coils with numbers of turns $N_1$ and $N_2$, respectively. A feedback loop controlled by the SQUID signal adjusts $I_2$ such that the ratio of the two resistors, to good approximation, equals the inverse ratio of the currents $I_1$ to $I_2$, which again is the inverse ratio of $N_1$ to $N_2$. A well-known current $kI_1$ is fed through an auxiliary winding $N_A = 1$ to improve the balance of the bridge, i.e., to bring the bridge voltage difference $\Delta U$, measured by a nanovoltmeter, close to zero. The unknown Hall resistance value of the QAHE device is calculated from the ratio between the measured remaining bridge voltage difference $\Delta U$ and the voltage drop $\Delta(IR)$ across each of the resistors. The approximate resistance ratio $R_K : 100\ \Omega$ is realized by choosing $N_1 = 16\ 004$ and $N_2 = 62$ (instead of $N_1 = 4130$ and $N_2 = 16$ used with the 12-bit CCC in earlier studies [12]). This results in increased magnetic fluxes (generated by $N_1$ and $N_2$) coupling into the SQUID, significantly reducing the SQUID contribution to the measurement uncertainty, particularly at very low current levels in the nA regime.

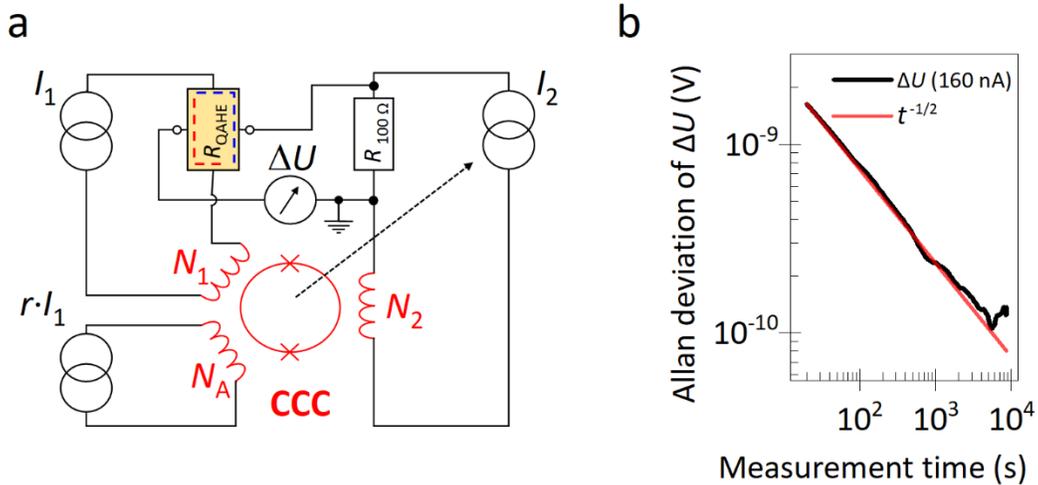

FIG. 2. **Schematic of the CCC setup and typical Allan deviation.** a) A simplified schematic of the CCC resistance bridge used to measure the Hall resistance of the QAHE device. The currents $I_1$ and $I_2$ are passed through the QAHE device and the 100 Ω reference resistor, respectively. The bridge is balanced by choosing turn numbers $N_1$ and $N_2$ in the coils to be equal to the ratio of the currents $I_2$ and $I_1$ (in first order) and by feeding an auxiliary winding $N_A$ with a well determined current $rI_1$. The SQUID-driven feedback loop (dashed black arrow) adjusts the current $I_2$ such that the total flux picked up from the windings $N_1$, $N_2$, and $N_A$ sensed by the SQUID is zero. The nanovoltmeter measures the voltage difference $\Delta U$, being close to zero when the bridge is balanced. b) Allan deviation plot of the bridge readings $\Delta U$ of a

5 h measurement on contact pair 2-3 with $I = 160$ nA. The curve follows a $t^{-1/2}$ relation, demonstrating the dominance of white noise up to at least about one hour of measurement (integration) time.

Current-dependent measurements are conducted at the dilution refrigerator base temperature of about 34 mK, at zero external magnetic field, and at a fixed gate voltage $U_{Gate} = 5.8$ V. The data for measurements on each contact pair is generated through 192 measurement cycles corresponding to a measurement time of 64 minutes, with currents ranging from 40 nA to 320 nA. A complete uncertainty budget with a more detailed discussion is provided in the "Supplementary Information". The Allan deviation of the bridge voltage $\Delta U$ is used to assess the stability of the measured CCC signal over time [20, 27, 28]. A typical Allan deviation plot is given in Fig. 2b, for a Hall resistance measurement at a current of 160 nA. It demonstrates that the Allan deviation of the bridge voltage signal follows a $t^{-1/2}$ relation up to about one hour, indicating the dominance of white noise during the typical measurement time, and thus justifies averaging the measurement data.

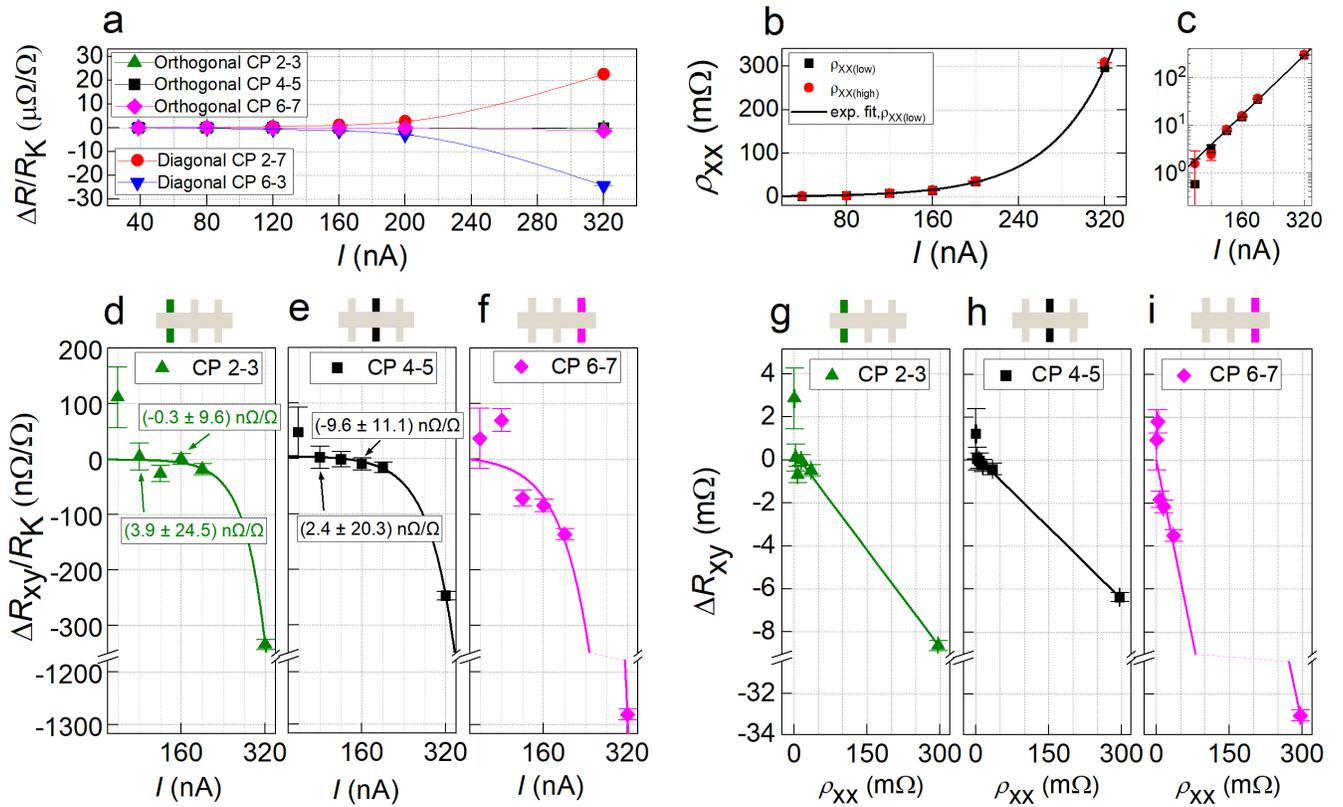

FIG. 3. **Precision measurement results.** Current-dependent CCC measurements for various contact pairs (abbreviated CP in the legends) at fixed gate voltage $U_{Gate} = 5.8$ V, temperature 34 mK and at zero external magnetic field. Each data point corresponds to 192 measurement cycles (in total 64 minutes). a) Relative deviation $\Delta R/R_K$ from $R_K$ for measurements on the three orthogonal contact pairs (2-3, 4-5, 6-7) and on the two diagonal contact pairs (2-7, 6-3), plotted as a function of the applied current. The colored lines are guides to the eye. b) Current dependence of the longitudinal resistivity $\rho_{xx}$ for the high (red color) and low (black color) potential edge. The black line represents an exponential fit (to $\rho_{xx(low)}$) following the relation $\rho_{xx} = \rho_{xx0} \exp(I/I_0)$, where $\rho_{xx0} = 915 \pm 59$ $\mu\Omega$ and $I_0 = 55.4 \pm 0.7$ nA. c) The same data plotted on a logarithmic $\rho_{xx}$ scale. d-f) Relative deviation of the Hall resistance from $R_K$ for the three orthogonal contact pairs, plotted as a function of the applied current. Solid lines are exponential fits following the relation $\Delta R_{xy}/R_K = a \exp(I/I_0) + \Delta R_{xy}/R_K (I = 0)$, where $\Delta R_{xy}/R_K (I = 0)$ is the deviation from quantization when extrapolated to zero current. g-i) Parametric plot of the Hall resistance deviation from quantization $\Delta R_{xy}$ for the same orthogonal contact pairs, plotted against the longitudinal resistivity $\rho_{xx}$ (along the low potential edge). Colored lines represent a linear fit for each contact pair, following the relation $\Delta R_{xy} = s\rho_{xx} + \Delta R_{xy}(\rho_{xx} = 0)$, where $s$ is the admixing parameter and $\Delta R_{xy}(\rho_{xx} = 0)$ is the deviation from Hall resistance quantization when extrapolated to zero longitudinal resistivity. All parameters for curves in d-i are listed in Table I.

Fig. 3a shows the relative deviation of the measured resistance signals from $R_K$, $\Delta R/R_K = (R - R_K)/R_K$, plotted as a function of the applied current for the three orthogonal contact pairs 2-3, 4-5, and 6-7, as well as the two diagonal contact pairs 2-7 and 6-3. A signal from the diagonal contact pairs represents a sum of Hall and longitudinal voltages. Therefore, the longitudinal voltages were determined by calculating the difference between the measured diagonal and orthogonal signals. The longitudinal resistance $R_{xx}$ along the device's high-potential (average of two configurations: CP23-CP63 and CP27-CP67) and low-potential (average of two configurations: CP27-CP23 and CP67-CP63) edges is then determined, and the longitudinal resistivity $\rho_{xx} = R_{xx}(W/L)$ for each of the high ($\rho_{xx(high)}$) and low potential edges ($\rho_{xx(low)}$), is calculated taking into account the aspect ratio of the device, where $W = 200~\mu$m and $L = 400~\mu$m.

The resulting $\rho_{xx}$ data are plotted in Fig. 3b as a function of the measurement current $I$. The data reveal an apparent exponential increase in the resistivity (solid line fit in the plot) as the applied current increases. $\rho_{xx}$ remains below about 20 mΩ for currents up to 160 nA (see Fig. 3c). Specifically, at a current of 80 nA we observe a value of $\rho_{xx} = (3.2 \pm 0.6)$ mΩ (on the low potential edge), which is slightly lower than the values reported previously: ≈ 6 mΩ in Ref. [21] (at $I = 1~\mu$A), and $(7.3 \pm 1.6)$ mΩ in Ref. [20] (at $I \leq 50$ nA). For currents higher than 160 nA, we observe a substantial increase in the longitudinal resistivity.

In Figs. 3d-f we focus on the relative deviation of the measured Hall resistance from $R_K$ measured on the orthogonal contact pairs 2-3, 4-5 and 6-7 as a function of applied current $I$. In the current range between 80 nA and 160 nA, the contact pair 4-5 exhibits Hall resistance quantization in agreement with $R_K$ within the measurement uncertainties (standard measurement uncertainties, coverage factor $k = 1$, are used throughout the paper). For a representative individual measurement at $I = 160$ nA on contact pair 2-3, $\Delta R_{xy}/R_K = (-0.3 \pm 9.6)$ nΩ/Ω at zero external magnetic field. When the measurement current is increased above about 200 nA, we observe a significant deviation of $\Delta R_{xy}/R_K$ from zero, which together with the significant increase in $\rho_{xx}$ represents the onset of the current-induced breakdown of QAHE [15, 16].

To further quantify the precision of QAHE quantization, we focus on the relationship between $\Delta R_{xy}$ and $\rho_{xx}$. This relationship holds particular significance in quantum Hall resistance metrology. In conventional GaAs-based QHE devices, a linear relationship $\Delta R_{xy} \sim s\rho_{xx}$ is usually observed [29–34], in which the parameter $s$ describes the admixing of the longitudinal resistance into the Hall resistance [22]. The value of $s$ can depend on the device geometry and operating conditions [2, 22, 29, 34–36]. Figs. 3g-i show parametric plots of the Hall resistance deviation from quantization, $\Delta R_{xy}$, for each orthogonal contact pair, plotted against the longitudinal resistivity $\rho_{xx}$ (along the low potential edge). To obtain the $s$-parameter value for each of the contact pairs, each data set is fitted with a linear relation (solid lines in Figs. 3g-i), $\Delta R_{xy} = s\rho_{xx} + \Delta R_{xy}(\rho_{xx} = 0)$, where $\Delta R_{xy}(\rho_{xx} = 0)$ is the deviation from quantization extrapolated to zero longitudinal resistivity. The resulting $s$-parameter value for each contact pair is listed in Table I.

| Contact pair | Extrapolation to zero current $I$. $\Delta R_{xy}/R_K(I) = \alpha \exp(I/I_0) + \Delta R_{xy}/R_K(I=0)$ | | | Extrapolation to zero longitudinal resistivity $\rho_{xx}$. $\Delta R_{xy}(\rho_{xx}) = s\rho_{xx} + \Delta R_{xy}(\rho_{xx} = 0)$ | |
|---|---|---|---|---|---|
| | $I_0$ [nA] | $\alpha$ [nΩ/Ω] | $\Delta R_{xy}/R_K(I=0)$ [nΩ/Ω] | $s$ [Ω/Ω] | $\Delta R_{xy}/R_K(\rho_{xx}=0)$ [nΩ/Ω] |
| 2-3 | 39.6 ± 22.7 | -0.10 ± 0.49 | **-1.3 ± 19.7** | -0.030 ± 0.002 | **13.2 ± 9.6** |
| 4-5 | 49.7 ± 10.5 | -0.40 ± 0.56 | **4.4 ± 8.7** | -0.022 ± 0.001 | **8.6 ± 6.7** |
| 6-7 | 56.1 ± 10.6 | -4.3 ± 4.8 | **5.1 ± 49.1** | -0.112 ± 0.003 | **0.5 ± 18.5** |

TABLE I. Values of the fitting parameters using both extrapolation methods (Fig. 3d-i), for all three orthogonal contact pairs.

The $s$-parameter values for contact pairs 2-3 and 4-5 are (-0.030 ± 0.002) and (-0.022 ± 0.001), respectively, which indicates an excellent geometrical alignment of both pairs of voltage contacts. The somewhat larger value (-0.112 ± 0.003) for the contact pair 6-7 could be a result of some local inhomogeneities in the device. Nevertheless, even a value of -0.1 is not uncommon for traditional QHRS, where this value typically ranges from -0.01 to as much as ~ -1 [2, 22, 29, 34–36]. Further systematic studies of more devices will be needed to determine the details behind the contact pair-to-pair variation in the $s$ parameter in magnetic TI samples.

Deviation from quantization as a result of longitudinal voltage admixing implies that a deviation from quantization $\Delta R_{xy}/R_K$ (Fig. 3d-f) is proportional to $\rho_{xx}$ (Fig. 3b-c). Considering that $\rho_{xx}$ is described remarkably well with an exponential function $\rho_{xx} = \rho_{xx0} \exp(I/I_0)$ (black curve in Figs. 3b-c), the $\Delta R_{xy}/R_K(I)$ signal can also be fitted using an exponential function model: $\Delta R_{xy}/R_K$

$= \alpha \exp(I/I_0) + \Delta R_{xy}/R_K$ ($I = 0$), where $\Delta R_{xy}/R_K$ ($I = 0$) is the deviation from quantization when extrapolated to zero current (the fitting parameters for each contact pair are listed in Table I). Note that, as expected, the value of $I_0$ for each contact pair fit is consistent within error bars with a value of $I_0 = 55.4\pm0.7$ nA from the $\rho_{xx}$ fitting in Fig. 3b.

Even the lowest recorded $\rho_{xx}$ values in magnetic TIs (a few mΩ) are considerably higher than the values typically recorded for the QHRS based on GaAs and graphene, which typically are of the order of a few $\mu\Omega$ [4, 37, 38]. This naturally implies that the effects of the voltage admixing on the quantization accuracy increase accordingly. In order to minimize this effect, we next analyze the relative deviation from quantization when the data is extrapolated using both methods: to zero current $\Delta R_{xy}/R_K(I = 0)$, and to zero longitudinal resistivity $\Delta R_{xy}/R_K$ ($\rho_{xx} = 0$) [29, 31, 33, 35–38], corresponding to ideal QHE conditions with vanishing dissipation. Using the fits from Figs. 3d-i, the extrapolated values for all three orthogonal contact pairs are listed in Table I. Each of the six fitting results shows a combined relative measurement uncertainty of less than 50 nΩ/Ω. The best result is obtained from the measurements at the center contact pair 4-5 (i.e., the pair of contacts furthest away from the hot spots in the corners of the device), with values of $\Delta R_{xy}/R_K$ ($\rho_{xx} = 0$) = (8.6 ± 6.7) nΩ/Ω and $\Delta R_{xy}/R_K$ ($I = 0$) = (4.4 ± 8.7) nΩ/Ω (each with combined standard uncertainty, $k = 1$), setting a new upper bound for the deviation of the Hall resistance in the QAHE regime from perfect quantization at zero external magnetic field. This improves the quantization precision at zero external magnetic field by about two orders of magnitude over the previous results: (170 ± 250) nΩ/Ω in Ref. [12], (-20 ± 310) nΩ/Ω in Ref. [20], about 1000 nΩ/Ω in Ref. [13], and about 2000 nΩ/Ω in Ref. [14]. Our results bring the precision of a zero external magnetic field QAHE on par with early conventional QHRS, establishing a zero external magnetic field quantum standard of resistance.

## Conclusions

We performed high-accuracy measurements of the Hall resistance quantization in the QAHE regime on a device based on a magnetic TI, $(V,Bi,Sb)_2Te_3$. When extrapolating to zero longitudinal resistivity (i.e., zero dissipation) or to zero current, the relative deviation of the Hall resistance from the von-Klitzing constant, measured at the center pair of Hall contacts in our device, is $\Delta R_{xy}/R_K$ ($\rho_{xx} = 0$) = (8.6 ± 6.7) nΩ/Ω and $\Delta R_{xy}/R_K$ ($I = 0$) = (4.4 ± 8.7) nΩ/Ω (each with combined standard uncertainty, $k = 1$) at zero external magnetic field, respectively, setting a new benchmark. The relative standard uncertainty for a typical individual data point at 160 nA for a measurement time of 64 minutes is ≤ 12 nΩ/Ω. Our results establish an operational quantum resistance standard at zero external magnetic field, which – if successfully integrated with the Josephson voltage standard – will provide an integrated universal quantum electrical reference.

## Acknowledgements

We gratefully acknowledge the financial support of the Free State of Bavaria (the Institute for Topological Insulators), Deutsche Forschungsgemeinschaft (SFB 1170, 258499086, and EXC-2123 QuantumFrontiers, 390837967), Würzburg-Dresden Cluster of Excellence on Complexity and Topology in Quantum Matter (EXC 2147, 39085490), the European Commission under the H2020 FETPROACT Grant TOCHA (824140), and by the project 23FUN07 QuAHMET, which has received funding from the European Partnership on Metrology, co-financed from the European Union's Horizon Europe Research and Innovation Programme and by the Participating States.

The authors gratefully acknowledge the support from Frank Hohls (PTB) with the experimental setup.

## Author Contributions

PTB: D. K. P. and M. Kr. performed the CCC measurements, together with the sample characterization. E. P. developed the 14-bit CCC used for the precision experiments. M. G. was involved in the development of the 14-bit CCC resistance bridge, in implementing the metrology-grade experiments to the PTB facilities, and in the analysis of measurement data. M. J. implemented the metrology-grade experiments at PTB. H. S. has contributed to the analysis of measurement data and supervised the research.

## Data Availability

All data that support the findings of this study are publicly available via PTB's Open Access Repositorium (PTB-OAR, https://oar.ptb.de/) under DOI: 10.7795/720.20240314.

# Zero External Magnetic Field Quantum Standard of Resistance at the $10^{-9}$ Level

D. K. Patel,[1,*] K. M. Fijalkowski,[2,3] M. Kruskopf,[1] N. Liu,[2,3] M. Götz,[1] E. Pesel,[1] M. Jaime,[1] M. Klement,[2,3] S. Schreyeck,[2,3] K. Brunner,[2,3] C. Gould,[2,3] L. W. Molenkamp,[2,3] and H. Scherer[1,†]

[1] *Physikalisch-Technische Bundesanstalt (PTB), Bundesallee 100, 38116 Braunschweig, Germany*
[2] *Faculty for Physics and Astronomy (EP3), Universität Würzburg, Am Hubland, D-97074, Würzburg, Germany*
[3] *Institute for Topological Insulators (ITI), Am Hubland, D-97074, Würzburg, Germany*

### GATE LEAKAGE TEST

Prior to the experiments, the top gate was tested for gate leakage. Fig. S1 depicts the leakage current characteristics of the device. The curves depicted by black and grey data points were measured as part of the basic characterization procedure of the device in a measurement setup based on lock-in amplifiers which are connected to the device via a breakout box and additional cables. The relative offset between the black curve (gate voltage sweep upwards) and the grey curve (gate voltage sweep downwards) is a consequence of the capacitive charging current, and a systematic shift from zero for both curves is an instrumental offset. The corresponding leakage resistance, including the cabling of this characterization setup, is 1.8 TΩ.

The curve depicted by the red data points was measured during the "precision measurement" phase with the CCC connected to the device. This avoids using the breakout box and coaxial cables. The corresponding leakage resistance here is 6.9 TΩ. The difference of about 5.1 TΩ between the gate leakage measurements obviously results from the differences in wiring leakage between the two setups.

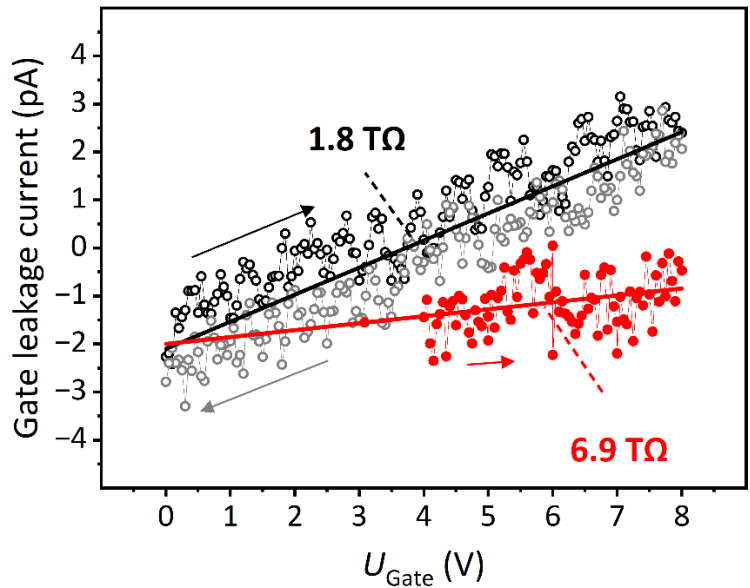

FIG. S1. **Results of gate leakage measurements.** Leakage currents as a function of the voltage $U_{\text{Gate}}$ applied to the top gate, measured at base temperature of about 34 mK. The measurements of the black and grey data points involve lock-in amplifiers connected to the device via a breakout box with additional cables, and the leakage resistance corresponding to the slope of the black line is 1.8 TΩ. The measurement of the red data points was performed without the breakout box and additional cables, and the leakage resistance corresponding to the slope of the red line is 6.9 TΩ.

## GATE VOLTAGE TUNING

Maximizing the insulating properties of the device's bulk is achieved by tuning the Fermi level via the gate voltage $U_{Gate}$. The optimal gate voltage minimizes the longitudinal resistance $R_{xx}$. Fig. S2 depicts results from measurements of $R_{xx}$ versus $U_{Gate}$ performed with a lock-in measurement at a frequency of 13.7 Hz and 5 nA (rms) of applied current. Since the limited measurement resolution at the base temperature (about 34 mK) does not allow for determining the minimum of $R_{xx}(U_{Gate})$, the measurement was repeated at an elevated temperature of 100 mK. From this, the gate voltage value $U_{Gate} = 5.8$ V was derived as the sweet spot for the QAHE (precision) measurements.

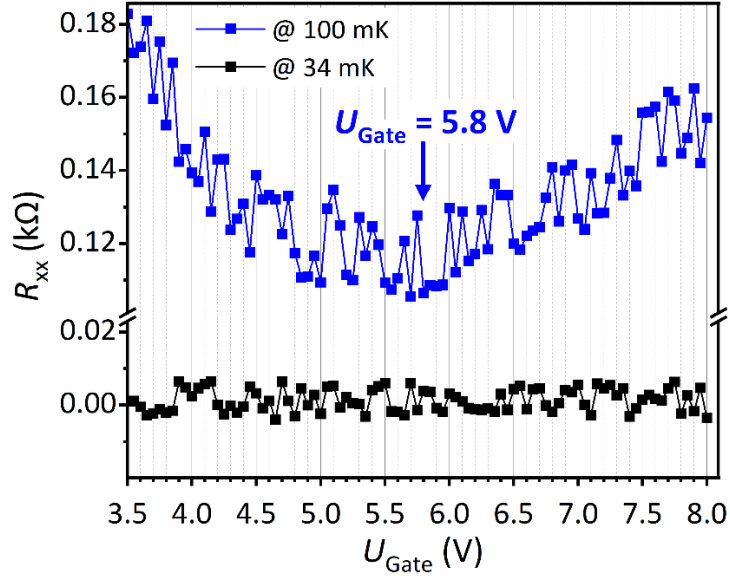

FIG. S2. **Results of a gate voltage tuning measurements.** Measured longitudinal resistance $R_{xx}$ versus a function of the voltage $U_{Gate}$ applied to the top gate. The optimal working point for the QAHE in terms of the gate voltage, identified as the minimum of $R_{xx}(U_{Gate})$ at $T = 100$ mK, is $U_{Gate} = 5.8$ V.

## REFERENCE RESISTOR

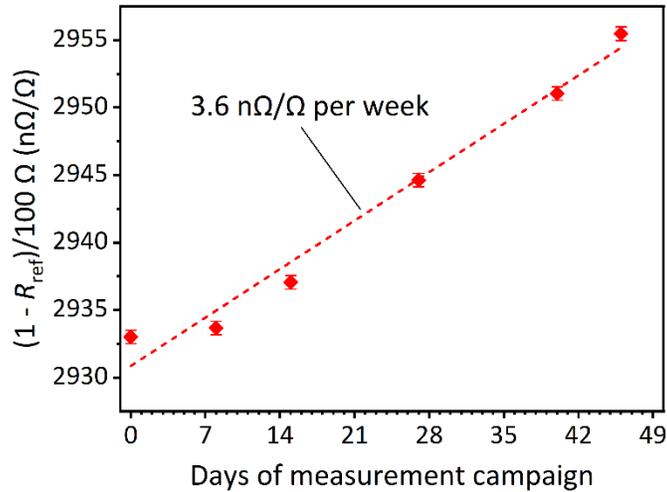

FIG. S3. **Drift behavior of the 100 Ω reference resistor during the measurement campaign.** The individual calibration values have uncertainties corresponding to error bars about the size of the diamond symbols.

The reference resistor used in our measurements has a nominal value of 100 Ω and is kept in a regulated thermostat chamber. Throughout the measurement campaign, its resistance was calibrated against the GaAs-based German resistance standard at PTB with calibration uncertainties below 0.5 nΩ/Ω. It showed a typical drift stability of 3.6 nΩ/Ω per week, as seen in Fig S3. As a conservative estimate for the uncertainty of the reference resistance, a contribution of 0.36 µΩ corresponding to the drift over one week is assigned in the uncertainty budget.

## UNCERTAINTY BUDGET ANALYSIS

We follow the internationally recommended "guide to the expression of uncertainty in measurement" (GUM) [S1] to estimate the uncertainty budget for the precision measurements performed with the cryogenic current comparator (CCC) resistance bridge (presented in section "Precision Measurements" in the main article). The measurement uncertainty budget given in Table S1 is evaluated from a typical measurement of the QAHE at a current level of 160 nA and using a 100 Ω reference resistor (cf. Fig. 2 in the main article). The unknown electrical resistance value $R_x$ is calculated according to

$$R_\mathrm{x} = R_\mathrm{ref} \cdot \frac{N_1 + r \cdot N_A}{N_2} \cdot \left(1 + \frac{\Delta U}{\Delta(I_2 R_\mathrm{ref})}\right) \cdot K_\mathrm{iso} \qquad (\text{Equ. 1})$$

where $R_\mathrm{ref}$ is the resistance of the reference resistor, $N_1$ and $N_2$ are the number of turns in the coils of the CCC, $r$ is a well-known factor describing the current injected into the auxiliary winding $N_A$, and $K_\mathrm{iso}$ is a factor quantifying the influence of imperfect insulation.

In the following, $u(x_i)$ is the standard uncertainty ($k = 1$) for the individual contributions of the quantities $x_i$. The sensitivity coefficients $c(x_i) = \partial R_X/\partial x_i$ are calculated from Equ. 1. With the abbreviated notations $P = V_{12}(1 + rV_{A1})$, $V_{12} = N_1/N_2$, $V_{A1} = N_A/N_1$, and $Q = 1 + \Delta U/\Delta(I_2 R_\mathrm{ref})$, the following relations for the sensitivity coefficients are derived:

$$c(R_\mathrm{ref}) = \frac{\partial R_\mathrm{x}}{\partial R_\mathrm{ref}} = P \cdot Q \cdot K_\mathrm{iso} \approx V_{12}$$

$$c(V_{12}) = \frac{\partial R_\mathrm{x}}{\partial V_{12}} = \frac{\partial R_\mathrm{x}}{\partial P} \cdot \frac{\partial P}{\partial V_{12}} = R_\mathrm{ref} \cdot Q \cdot K_\mathrm{iso} \cdot (r \cdot V_{A1} + 1) \approx R_\mathrm{ref}$$

$$c(V_{A1}) = \frac{\partial R_\mathrm{x}}{\partial V_{A1}} = \frac{\partial R_\mathrm{x}}{\partial P} \cdot \frac{\partial P}{\partial V_{A1}} = R_\mathrm{ref} \cdot Q \cdot K_\mathrm{iso} \cdot r \cdot V_{12} \approx R_\mathrm{ref} \cdot r \cdot V_{12}$$

$$c(r) = \frac{\partial R_\mathrm{x}}{\partial r} = \frac{\partial R_\mathrm{x}}{\partial P} \cdot \frac{\partial P}{\partial r} = R_\mathrm{ref} \cdot Q \cdot K_\mathrm{iso} \cdot V_{12} \cdot V_{A1} \approx R_\mathrm{ref} \cdot V_{12} \cdot V_{A1}$$

$$c(\Delta U) = \frac{\partial R_\mathrm{x}}{\partial(\Delta U)} = \frac{\partial R_\mathrm{x}}{\partial Q} \cdot \frac{\partial Q}{\partial(\Delta U)} = \frac{R_\mathrm{ref} \cdot P \cdot K_\mathrm{iso}}{\Delta(I_2 R_\mathrm{ref})} \approx \frac{R_\mathrm{ref} \cdot V_{12}}{\Delta(I_2 R_\mathrm{ref})}$$

$$c(\Delta(I_2 R_\mathrm{ref})) = \frac{\partial R_\mathrm{x}}{\partial(\Delta(I_2 R_\mathrm{ref}))} = \frac{\partial R_\mathrm{x}}{\partial Q} \cdot \frac{\partial Q}{\partial(\Delta(I_2 R_\mathrm{ref}))} = -\frac{R_\mathrm{ref} \cdot P \cdot K_\mathrm{iso} \cdot \Delta U}{(\Delta(I_2 R_\mathrm{ref}))^2} \approx -\frac{R_\mathrm{ref} \cdot V_{12} \cdot \Delta U}{(\Delta(I_2 R_\mathrm{ref}))^2}$$

$$c(K_\mathrm{iso}) = \frac{\partial R_\mathrm{x}}{\partial K_\mathrm{iso}} = R_\mathrm{ref} \cdot P \cdot Q \approx R_\mathrm{ref} \cdot V_{12}$$

The uncertainty contributions are computed as $u_i = u(x_i) \cdot c(x_i)$, with $u(x_i) = s(x_i)/\Gamma_\mathrm{PDF}$, where $s(x_i)$ are the standard errors of the quantities $x_i$, and $\Gamma_\mathrm{PDF}$ is the normal probability distribution function: $\Gamma_\mathrm{PDF} = 1$ for type A (i.e., stochastic) contributions, and $\sqrt{3}$ for type B (i.e., systematic) contributions with a rectangular distribution.

The determination of the standard errors $s(x_i)$ for the individual uncertainty contributions was performed analogously to Ref. [S2], giving the following figures:

$\Delta U$ is the type A uncertainty contribution for the bridge voltage measurements over 64 minutes, contributing with 0.14 nV in the following example.

$\Delta U$(Instrum.) is the type B component for the uncertainty contributions resulting from the bridge amplifier's gain and the subsequent voltage-frequency-voltage conversion chain, contributing with about 0.06 nV.

$\Delta(IR)$ is the type B component for the uncertainty contribution measured with a multimeter, contributing with about 8.3 µV.

"Gate leakage" can inject extra current through the gate insulation of the device. Since voltage applied to the gate electrode is kept constant throughout the measurements, to first order, the leakage current is constant. However, as the measurement bias current is periodically reversed, the edge state voltage is periodically changing between zero and $U_H$ (Hall voltage). Therefore, a voltage drop over the gate dielectric is periodically changing with an amplitude equal to $U_H$, resulting in periodic fluctuations in gate leakage current with an amplitude equal to $U_H$ divided by the insulation resistance ($R_{iso}$). This second-order effect, being in sync with the bias current reversals, can have an influence on the measured $\Delta U$. A conservative upper bound on the size of this effect is given by a ratio of the periodic leakage current amplitude ($U_H/R_{iso}$) and the measurement bias current. This ratio is equal to $R_K/R_{iso}$, and taking a lower bound of 6.9 TΩ for $R_{iso}$ (as measured in this setup), the corresponding conservative estimate of type B uncertainty contribution is taken as 25813 Ω / 6.9 TΩ = 3.7 $10^{-9}$.

$R_{ref}$ is the type B component for the calibration uncertainty contribution of the 100 Ω reference resistor, according to the section "Reference Resistor" contributing with 0.36 µΩ.

$\Delta U$(SQUID) is the type B component for the uncertainty contribution from the SQUID non-linearity, which for the 14-bit CCC contributes about 0.009 nV.

$N_1/N_2$ is the type B component for the uncertainty contribution for the numbers-of-turns ratio of the coils in the primary and secondary branches of the 14-bit CCC, contributing about 2.6 $10^{-8}$.

$K_{iso}$ describes the influence of imperfect insulation of the wiring connecting the QAHE device. A deviation from $K_{iso} = 1$ (for the case of perfect insulation) is accounted for as a type B uncertainty contribution derived from the relevant insulation resistance, which was experimentally determined to be at least 10 TΩ in our setup. For the case of our setup, being grounded on the CCC branch of the 100 Ω reference resistor (cf. Fig. 2a of the main article), the corresponding uncertainty contribution is calculated from the ratio of the reference resistor and the lower bound of the insulation resistance, i.e., 100 Ω / 10 TΩ = 1 $10^{-11}$.

Note that the auxiliary coil ($N_A$ according to Fig. 2a of the main article) was not used for the CCC measurements, so that uncertainty contributions for the numbers-of-turns ratio ($N_A/N_1$) and for the current injected into the auxiliary winding ($r$) are zero and suppressed in the following uncertainty budget.

The uncertainty budget provided in Table S1 is dominated by the type A uncertainty for the bridge voltage $\Delta U$, followed by the two type B uncertainty contributions $\Delta U$(Instrum.) and $\Delta(IR)$. The total combined relative standard uncertainty ($k = 1$) for this specific case is ≤ 12 nΩ/Ω.

| Representative error budget (for an individual 160 nA measurement on contact pair 6-7) | | | | | | |
|---|---|---|---|---|---|---|
| Type | $x_i$ | $s(x_i)$ | $\Gamma_{PDF}$ | $u(x_i)$ | $|c(x_i)|$ | $|u_i|$ |
| A | $\Delta U$ | 8.13 $10^{-11}$ V | 1 | 8.13 $10^{-11}$ V | 3.13 $10^6$ A$^{-1}$ | 0.255 mΩ |
| B | $\Delta U$(Instrum.) | 5.58 $10^{-11}$ V | $\sqrt{3}$ | 3.22 $10^{-11}$ V | 3.14 $10^6$ A$^{-1}$ | 0.101 mΩ |
| B | $\Delta(IR)$ | 8.26 $10^{-6}$ V | $\sqrt{3}$ | 4.77 $10^{-6}$ V | 21.07 A$^{-1}$ | 0.101 mΩ |
| B | Gate leakage | 3.7 $10^{-9}$ | $\sqrt{3}$ | 2.14 $10^{-9}$ | 2.58 $10^4$ Ω | 55.14 µΩ |
| B | $R_{ref}$ | 3.6 $10^{-7}$ Ω | $\sqrt{3}$ | 2.08 $10^{-7}$ Ω | 2.58 $10^2$ | 53.65 µΩ |
| B | $\Delta U$(SQUID) | 8.96 $10^{-12}$ V | $\sqrt{3}$ | 5.17 $10^{-12}$ V | 3.13 $10^6$ A$^{-1}$ | 16.19 µΩ |
| B | $N_1/N_2$ | 2.58 $10^{-8}$ | $\sqrt{3}$ | 1.49 $10^{-8}$ | 100 Ω | 1.49 µΩ |
| B | $K_{iso}$ | 1.0 $10^{-11}$ | $\sqrt{3}$ | 5.77 $10^{-12}$ | 2.58 $10^4$ Ω | 0.15 µΩ |
| | | Root sum of squares $\sqrt{\Sigma_i u_i^2}$ = 0.302 mΩ | | | | |
| | | Combined relative measurement uncertainty = 0.302 mΩ / 25812.807459 Ω = **11.7 nΩ/Ω** | | | | |

Table S1. Measurement uncertainty budget in the case of a current level of 160 nA and a measurement time of 64 minutes (contact pair 6-7). $u_i$ are the individual standard uncertainty contributions with corresponding sensitivity coefficients $c(x_i)$.